\documentclass[aps,prb,preprint,showpacs,showkeys,floatfix]{revtex4}

\usepackage{graphicx,color}
\usepackage{multirow,slashbox}

\newcommand{\rev}[1]{\textcolor{black}{#1}}

\graphicspath{{figs/}}
\begin{document}

\title{Misfit strain-induced buckling for transition-metal dichalcogenide lateral heterostructures: a molecular dynamics study}

\author{Jin-Wu Jiang}
    \altaffiliation{Corresponding author: jiangjinwu@shu.edu.cn; jwjiang5918@hotmail.com}
    \affiliation{Shanghai Institute of Applied Mathematics and Mechanics, Shanghai Key Laboratory of Mechanics in Energy Engineering, Shanghai University, Shanghai 200072, People's Republic of China}

\date{\today}
\begin{abstract}

Molecular dynamics simulations are performed to investigate the misfit strain-induced buckling of the transition-metal dichalcogenide (TMD) lateral heterostructures, denoted by the seamless epitaxial growth of different TMDs along the in-plane direction. The Stillinger-Weber potential is utilized to describe both the interaction for each TMD and the coupling between different TMDs, i.e., MX$_2$ (with M=Mo, W and X=S, Se, Te). It is found that the misfit strain can induce strong buckling of the free-standing TMD lateral heterostructures of large area, resulting from the TMDs' atomic-thick nature. The buckling phenomenon occurs in a variety of TMD lateral heterostructures of different compositions and in various patterns. Our findings raise a fundamental mechanical challenge for the structural stability of the free-standing  TMD lateral heterostructures.

\end{abstract}
\maketitle
\pagebreak

\section{Introduction}

The layered transition-metal dichalcogenides (TMDs) have attracted considerable attention as semiconductors with tunable electronic and other distinctive electro-optical properties. The growth of TMDs in the form of lateral heterostructures is a powerful technique to construct lateral p-n diodes with excellent current rectification and other functional devices based on the TMDs. In 2014, several experiments successfully fabricated the TMD lateral heterostructures and illustrated some novel properties. The WSe$_2$-MoSe$_2$ lateral heterostructure was produced by using a mixture of WSe$_2$ and MoSe$_2$ powders as the source for the chemical vapour deposition (CVD) process.\cite{HuangC2014nm} The WS$_2$-MoS$_2$ lateral heterostructure was synthesized by controling the growth temperature, where atomic visualization discloses the preference of the lateral interface along the zigzag direction.\cite{GongY2014nm} Another experiment also revealed the zigzag direction to be the preferable growth direction for the lateral interface.\cite{ZhangXQ2015nl} Duan et al. synthesized the MoS$_2$-MoSe$_2$ and WS$_2$-WSe$_2$ lateral heterostructures by \textit{in situ} modulation of the reactants during the CVD growth process, where the synthetic conditions are kept unchanged during the switch of the reactants, so that the edge is fresh and unpassivated and can serve as the active growth front for successive lateral epitaxial growth.\cite{DuanX2014nn}

In the past few years, considerable efforts have been made to improve the growth of the TMD lateral heterostructure. Various modified CVD techniques were introduced to synthesize TMD lateral heterostructures, aiming mainly at more precise control over patterns and sharper lateral interface.\cite{ChenK2015acsn,ChenK2015adm,GongY2015nl,ChenJ2016cm,LingX2016adm,LiuB2016acsn,SonY2016nl,ChenX2017acsami,LiMY2018afm,SahooPK2018nat} Li et al. developed a two-step epitaxial growth of the seamless WSe$_2$-MoS$_2$ lateral heterostructure (of different metals and chalcogen) with atomically sharp lateral interfaces, through proper control over the CVD conditions to avoid possible alloy reactions.\cite{LiMY2015sci} Recently, Zhang et al. designed a robust step-by-step CVD process to grow diverse TMD lateral heterostructures with multiple distinct TMD blocks, by preventing undesired thermal degradation and uncontrolled homogeneous nucleation.\cite{ZhangZ2017sci} Experiments have revealed some important mechanisms that can affect the synthesis of the TMD lateral heterostructures, such as the diffusion mechanism\cite{BogaertK2016nl} and the layer number.\cite{LiH2017acsn} Though TMD lateral heterostructures have been mostly grown by the CVD method, the lithography technique has been used to assist the creation of TMD lateral heterostructures in predefined patterns,\cite{Mahjouri-SamaniM2015nc,LiH2016acsn} and the pulsed-laser-deposition-assisted selenization method was used to grow the MoSe$_2$-WSe$_2$ lateral heterostructure.\cite{UllahF2017acsn} Besides the two-phase heterostructures, some other TMD lateral structures with tunable properties have also been synthesized in recent years, such as the TMD alloys with randomly distributed composition (eg. Mo$_{1-x}$W$_{x}$S$_2$)\cite{KobayashiY2015nnr,ZhangW2015nns,YoshidaS2015sr,DuanX2016nl,LiuX2017afm,AslanOB2018nl,ApteA2018acsn} or spatially graded composition.\cite{LiH2015jacs,ZhengS2015apl,WuX2017nns,LiZ2017acsami}

As a characteristic feature of the coherent interface, the misfit strain plays an important role in physical properties of the interface in TMD lateral heterostructures. By growing TMD coherent lateral heterostructures with controlled supercell dimensions, Xie et al. illustrated the tunability of optical properties by the misfit strain.\cite{XieS2018sci} Zhang et al. measured the distribution of the misfit strain tensor by examining the Moire pattern and investigated the misfit strain effect on the band alignment of the WSe$_2$-MoS$_2$ lateral heterostructures.\cite{ZhangC2018nn} From the mechanics point of view, TMDs are atomic membranes of high flexibility, so these structures are easy to buckle under compression. The misfit will cause compressive stress in the TMD of larger lattice constant in the lateral heterostructure, where the buckling instability may take place. This misfit strain-induced buckling has been discussed for the graphene and boron nitride lateral heterostructure.\cite{AlredJM2015nr} The possibility of this misfit strain-induced buckling is thus one of the topic discussed in the present work.

Although experiments have gained significant progress on synthesizing TMD lateral heterostructures, present theoretical studies are quite limited and mostly focus on the electronic band alignment for TMD alloys\cite{LiuX2017afm,ShiZ2018acsaem} or TMD lateral heterostructures,\cite{WeiW2015sr,WeiW2015pccp,GuoY2016apl,WeiW2016pccp,ZhangJ2017tdm,WeiW2017pccp,ArasM2018jpcc} the electronic transport properties,\cite{KangJ2015jpcc,AnY2016jmcc,YangZ2017pccp,CaoZ2017jpcl} or some possible applications.\cite{SunJ2016rscadv,YangY2017jpcl} Few studies investigated the possibility of designing lateral heterostructres by the first-principles calculations.\cite{SunQ2016tdm,JinH2016jmcc,LeenaertsO2016tdm,LeenaertsO2017apl,ChengK2017jmcc,SunQ2017tdm,SunQ2017nnr,YuanJ2018ass} All of these theoretical studies are based on the first-principles calculations approach, which is accurate but of expensive computational costs. Theoretical investigations are needed for many emerging mechanical or thermal properties of the TMD lateral heterostructures, such as the distribution of the misfit strain, the thermal transport across the lateral interface, the sharpness of the lateral interface, etc. These properties are sensitive to the size of the structure, and typically require more than tens of thousands of atoms, which is out of the capability of first-principles calculations. An empirical potential is thus highly desirable for these investigations, which will be developed in the present work.

In this work, we perform molecular dynamics simulations to study the misfit strain-induced buckling in the TMD lateral heterostructures. The Stillinger-Weber (SW) potential is used to describe both the interaction within each TMD and the coupling between different TMDs. We calculate the strain distribution within the TMD lateral heterostructure, and the results are consistent with the experiment. Misfit strain can cause the buckling of TMD lateral heterostructures with the size larger than a critical value, due to the atomic-thick nature of the TMD layers. The misfit strain-induced buckling is found to be a general phenomenon in TMD lateral heterostructures constructed by different TMD constitutions or of different patterns.

\section{Simulation details}

The SW potential is one of the most efficient nonlinear empirical potentials, and will be used to describe the interaction for TMD lateral heterostructures in the present work. It contains the following two-body and three-body interaction terms,
\begin{eqnarray}
V_{2}\left(r_{ij}\right) & = & \epsilon A\left(B\sigma^{p}r_{ij}^{-p}-\sigma^{q}r_{ij}^{-q}\right)e^{[\sigma\left(r_{ij}-a\sigma\right)^{-1}]}\\
\label{eq_sw2}
V_{3}\left(\vec{r}_{i},\vec{r}_{j},\vec{r}_{k}\right) & = & \epsilon\lambda e^{\left[\gamma\sigma\left(r_{ij}-a\sigma\right)^{-1}+\gamma\sigma\left(r_{jk}-a\sigma\right)^{-1}\right]}\left(\cos\theta_{jik}-\cos\theta_{0}\right)^{2}
\label{eq_sw3}
\end{eqnarray}
where $r_{ij}$ is the distance between atoms i and j, and $\theta_{jik}$ is the angle formed by bonds $r_{ji}$ and $r_{jk}$. $\theta_0$ is the equilibrium bond angle. Other quantities are SW potential parameters to be parametrized for different materials. We have recently parameterized the SW potential for most atomic-thick-layered materials including individual TMDs, i.e., MX$_2$ (with M=Mo, W and X = S, Se, Te).\cite{JiangJW2017intech}

There are two inequivalent bond angles in the hexagonal MX$_2$, which, however, are very close to each other for MX$_2$ with M=Mo, W and X = S, Se, Te. For example, these two inequivalent angles are $82.119^{\circ}$ and $81.343^{\circ}$ in MoSe$_2$.\cite{JiangJW2017intech} We can thus simplify the configuration of the MX$_2$ by using a single bond angle $\theta_0=80.581^{\circ}$. This simplification gives reasonably accurate results for MoS$_2$.\cite{JiangJW2015sw} The simplification is applicable to MX$_2$ with M=Mo, W and X=S, Se, Te, and is adopted in the present work.

\begin{figure}[tb]
  \begin{center}
    \scalebox{1}[1]{\includegraphics[width=8cm]{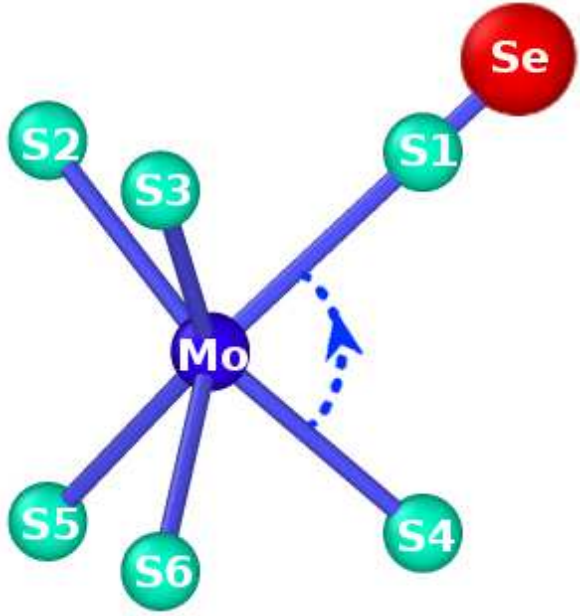}}
  \end{center}
  \caption{(Color online) Local environment around atom Mo in the MoS$_2$ hexagonal lattice with the Mo layer sandwiched by two S layers. Atoms S$_1$, S$_2$, and S$_3$ are in the top group of the MoS$_2$ layer. Atoms S$_4$, S$_5$, and S$_6$ are in the bottom group of the MoS$_2$ layer. Atom S$_1$ is substituted by atom Se. Note that the bending of angles like $\theta_{\rm MoS_1S_6}$ is not considered in the SW potential, and a modification is necessary for LAMMPS to exclude this term in the three-body SW potential (see the appendix).}
  \label{fig_cfg_6atom}
\end{figure}

In the TMD lateral heterostructures, there are crossing interactions between two TMDs, where new terms of the three-body form in Eq.~(\ref{eq_sw3}) are needed. Let's take MoS$_2$-WSe$_2$ lateral heterostructure as an example. At the interface, there are crossing angles like $\theta_{\rm MoSSe}$, which can be regarded as the substitution of one S atom by the Se atom in the angle $\theta_{\rm MoSS}$ as illustrated in Fig.~\ref{fig_cfg_6atom}. We focus on the angle $\theta_{\rm MoS_1S_4}$ (with Mo as the apex). The substitution of S$_1$ atom by Se atom in Fig.~\ref{fig_cfg_6atom} can cause two competing effects on the equilibrium value of angle $\theta_{\rm MoS_1S_4}=\theta_0$. First, Se has a larger nuclear radius than S (i.e., Se has more electrons than S), so the repulsive force between Se-S$_4$ is stronger than S$_1$-S$_4$, which shall enlarge the equilibrium angle $\theta_0$. Second, the Se atom is further away from the Mo atom (i.e. Mo-Se bond is longer than Mo-S$_1$ bond), which tends to reduce the value of angle $\theta_0$. These two competing effects cancel with each other, resulting in only neglectable changes of the equilibrium angle $\theta_0$. It is thus reasonable to assume that the equilibrium angle $\theta_0=80.581^{\circ}$ keeps unchanged in the crossing angles like $\theta_{\rm MoSeS_4}$. We thus use $\theta_0=80.581^{\circ}$ for all bond angles in the TMD lateral heterostructures. It should be noted that the bending of angles like $\theta_{\rm MoS_1S_6}$ is not considered in the SW potential, and a modification is necessary for LAMMPS to exclude this term in the three-body SW potential (see the appendix).

The energy parameter $\lambda_{\rm MXX'}$ in Eq.~(\ref{eq_sw3}) for the crossing angle $\theta_{\rm MXX'}$ is derived by the geometric average,
\begin{eqnarray}
\lambda_{\rm MXX'} & = & \sqrt{\lambda_{\rm MXX}\lambda_{\rm MX'X'}}
\label{eq_lambda}
\end{eqnarray}
where $\lambda_{\rm MXX}$ and $\lambda_{\rm MX'X'}$ are angle bending parameters corresponding to $\theta_{\rm MXX}$ and $\theta_{\rm MX'X'}$ in the individual TMD, respectively.\cite{JiangJW2017intech} There is no new term of the two-body form. We have generated all necessary terms for the SW potential of the multicomponent MX$_2$ system with M=Mo, W and X=S, Se, Te. The SW potential script for LAMMPS can be found from our group website http://jiangjinwu.org/sw. Specifically, this SW potential script is applicable to various TMD compounds, such as alloys and lateral heterostructures. We note that the largest system simulated in the present work contains more than 50,000 atoms, which goes beyond the computational capability of first-principles approaches and thus exhibits the power of the SW empirical potential.

Molecular dynamics simulations are performed to simulate the TMD lateral heterostructures. The TMD lateral heterostructures are first relaxed to the energy-minimum configuration by the conjugate gradient (CG) algorithm. The relaxed structure is further thermalized within the NPT (constant particle number, constant pressure, and constant temperature) ensemble for 100~ps by the Nos\'e-Hoover approach.\cite{Nose,Hoover} \rev{A low temperature of 4.2~K is used, so that the thermal expansion effect is neglectable and the buckling is intrinsically caused by the misfit strain. The thermal vibration becomes stronger at higher temperature, which shall increase the possibility of buckling.} The standard Newton equations of motion are integrated in time using the velocity Verlet algorithm with a time step of 1~{fs}. Simulations are performed using the publicly available simulation code LAMMPS~\cite{PlimptonSJ}, while the OVITO package is used for visualization~\cite{ovito}.

\section{Results and discussions}

\begin{figure*}[tb]
  \begin{center}
    \scalebox{1}[1]{\includegraphics[width=\textwidth]{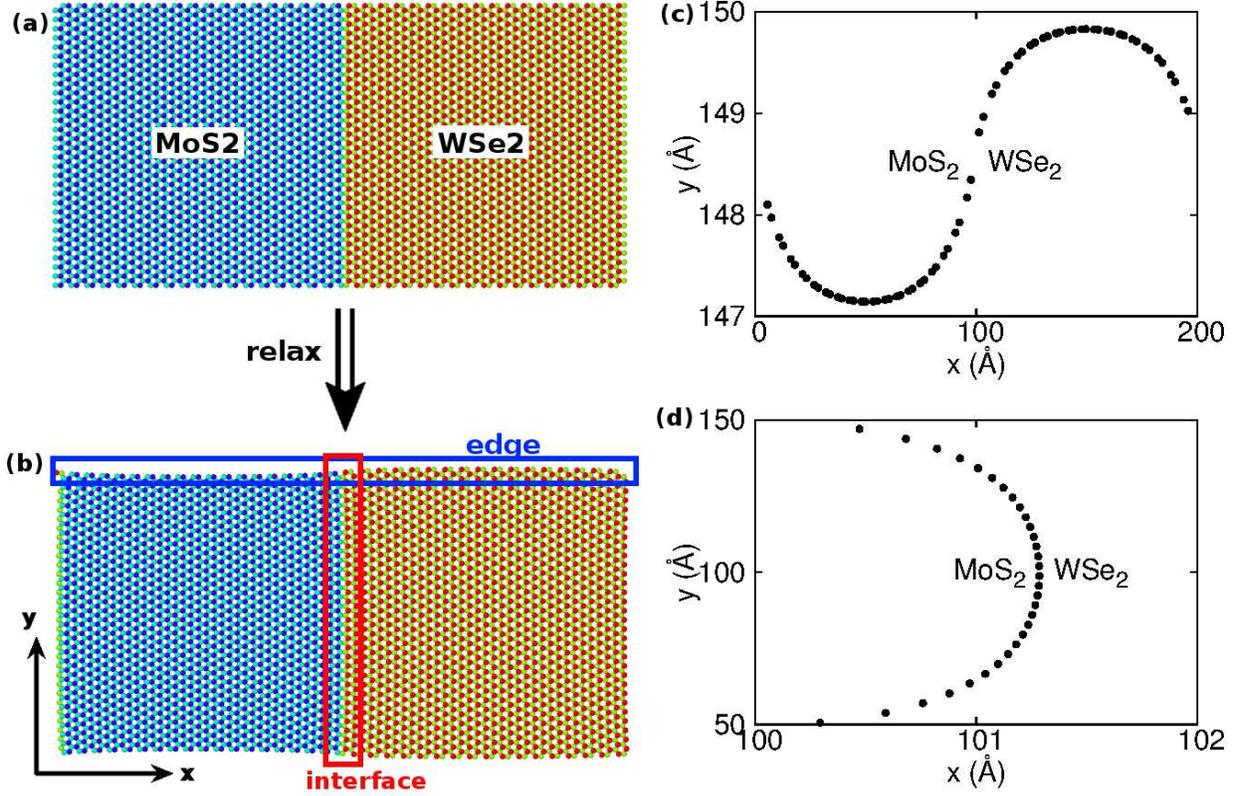}}
  \end{center}
  \caption{(Color online) Configuration of the MoS$_2$-WSe$_2$ lateral heterostructure of dimension $L_x\times L_y =200\times 100$~{\AA}. (a) Ideal unrelaxed structure. (b) Relaxed structure. The top edge (blue box) and the interface (red box) are deformed. (c) Details for the deformation of the top edge. (d) Details for the deformed interface illustrate the bending of the interface with MoS$_2$ (WSe$_2$) on the inner (outer) side.}
  \label{fig_cfg}
\end{figure*}

\begin{figure}[tb]
  \begin{center}
    \scalebox{1}[1]{\includegraphics[width=8cm]{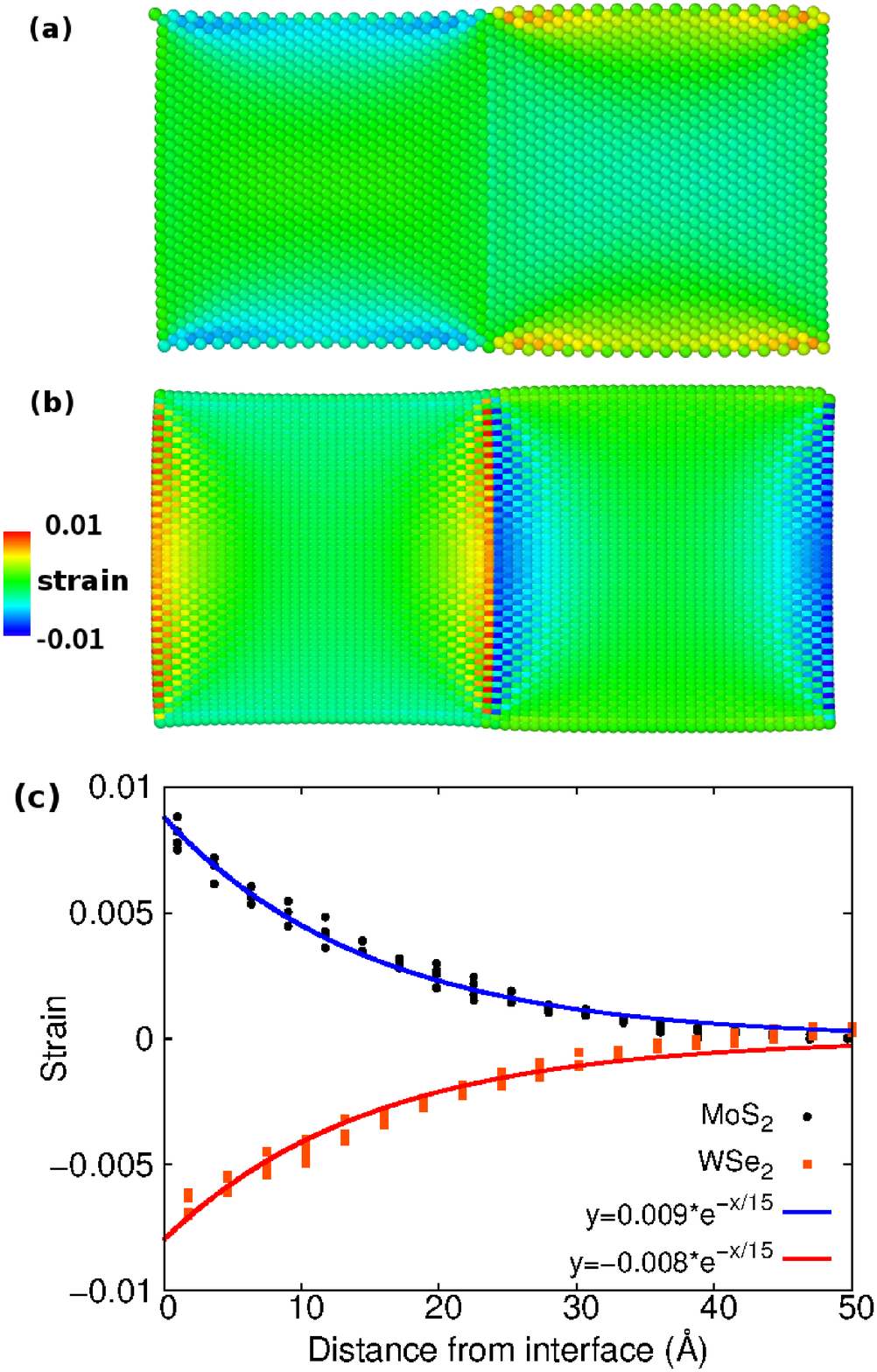}}
  \end{center}
  \caption{(Color online) Strain distribution within the relaxed configuration of MoS$_2$-WSe$_2$ lateral heterostructure of dimension $L_x\times L_y =200\times 100$~{\AA}. (a) Strain for bonds perpendicular to the interface. (b) Strain for bonds not perpendicular to the interface. (d) Strain versus the distance away from the central interface for atoms with $y=L_y/2$.}
  \label{fig_strain}
\end{figure}

Fig.~\ref{fig_cfg} shows the configuration for the MoS$_2$-WSe$_2$ lateral heterostructure of dimension $L_x\times L_y=200\times 100$~{\AA}. \rev{The dimension in the z direction (i.e., thickness) is not well defined in the atomic-thick nanomaterials. The thickness can be defined either as the space between two layers in the bulk system or as the size of the atom. However, this quantity is not used in the calculations of the present work, so we will not give an exact value for this quantity.} MoS$_2$ is on the left while WSe$_2$ is on the right, forming a vertical interface in the middle. The interface is of the preferred zigzag shape as observed in the experiment.\cite{GongY2014nm,ZhangXQ2015nl} Fig.~\ref{fig_cfg}~(a) shows the ideal configuration, where MoS$_2$ and WSe$_2$ are constructed with a common lattice constant. The ideal configuration is relaxed to the energy-minimum state as shown in Fig.~\ref{fig_cfg}~(b). Periodic boundary condition is applied in the x-direction, and free boundary condition is applied in the y- and z-directions. The lattice constant of WSe$_2$ is larger than MoS$_2$, resulting in obvious misfit strain at the interface. The heterostructure is deformed by the misfit strain. Fig.~\ref{fig_cfg}~(c) is the enclosed configuration for the top edge. MoS$_2$ is stretched at the left side of interface, while WSe$_2$ is compressed at the right side of the interface, leading to the compressive (tensile) stress on the left (right) side of the interface. As a result, the interface is slightly bent with MoS$_2$ (WSe$_2$) on the inner (outer) side as shown in Fig.~\ref{fig_cfg}~(d).

We further examine the strain distribution in Fig.~\ref{fig_strain} for the MoS$_2$-WSe$_2$ lateral heterostructure of dimension $L_x\times L_y=200\times 100$~{\AA}. Fig.~\ref{fig_strain}~(a) shows the strain distribution for bonds perpendicular to the interface. For the MoS$_2$ on the left of the interface, there is only small tensile strain in the interior region, and the top and bottom edges of the MoS$_2$ region are compressed due to the bending of the interface displayed in Fig.~\ref{fig_cfg}~(d) (with MoS$_2$ on the inner side). For the WSe$_2$ on the right of the interface, there is small compressive strain in the interior region, and there are some tensile strains at the top and bottom edges due to the bending of the interface displayed in Fig.~\ref{fig_cfg}~(d) (with WSe$_2$ on the outer side). Fig.~\ref{fig_strain}~(b) illustrates the strain distribution for these bonds that are not perpendicular to the interface. Large strains are found at the middle interface. There is another interface formed by the left and right ends, due to the periodic boundary condition applied in the x-direction. We will focus on the interface at the middle of the system, i.e., at $x=L_x/2$. The strain decays rapidly away from the interface. Fig.~\ref{fig_strain}~(c) is the strain versus distance from the interface for atoms at $y=L_y/2$. Both strains in MoS$_2$ (tensile) and WSe$_2$ (compressive) decays exponentially with a critical length of 15.0~{\AA}. The exponential decay of the strain was also observed in a recent experiment.\cite{ZhangC2018nn} The strain for bonds perpendicular to the interface is much smaller than the strain for bonds not perpendicular to the interface, which is consistent with the experiment.\cite{XieS2018sci}

\begin{figure}[tb]
  \begin{center}
    \scalebox{1}[1]{\includegraphics[width=8cm]{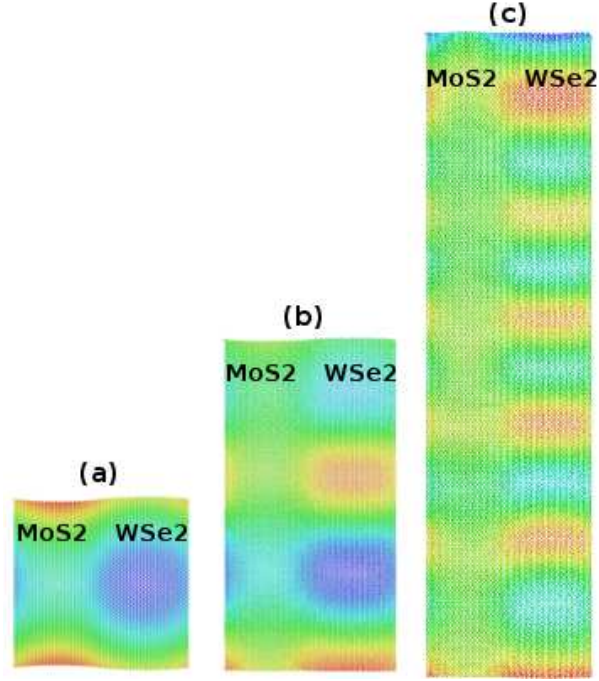}}
  \end{center}
  \caption{(Color online) Buckling of the MoS$_2$-WSe$_2$ lateral heterostructure of length $L_x=200$~{\AA} at 4.2~K. The width is $L_y=$200, 400, 800~{\AA} for the structures shown in (a), (b), and (c), respectively. The WSe$_2$ in the right part buckles. Color represents the atomic z-coordinate.}
  \label{fig_buckling_mos2-wse2}
\end{figure}

\begin{figure}[htpb]
  \begin{center}
    \scalebox{1}[1]{\includegraphics[width=8cm]{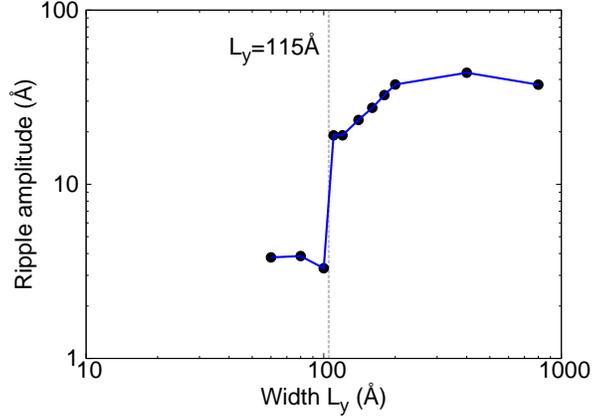}}
  \end{center}
  \caption{(Color online) The amplitude of the buckling ripple for MoS$_2$-WSe$_2$ lateral heterostructure of different width $L_y$ at 4.2~K. The length is fixed at $L_x=200$~{\AA}. The vertical grey line locates the critical width $L_y=105$~{\AA}, above which buckling occurs.}
  \label{fig_buckling_amplitude}
\end{figure}

\begin{figure}[htpb]
  \begin{center}
    \scalebox{1}[1]{\includegraphics[width=8cm]{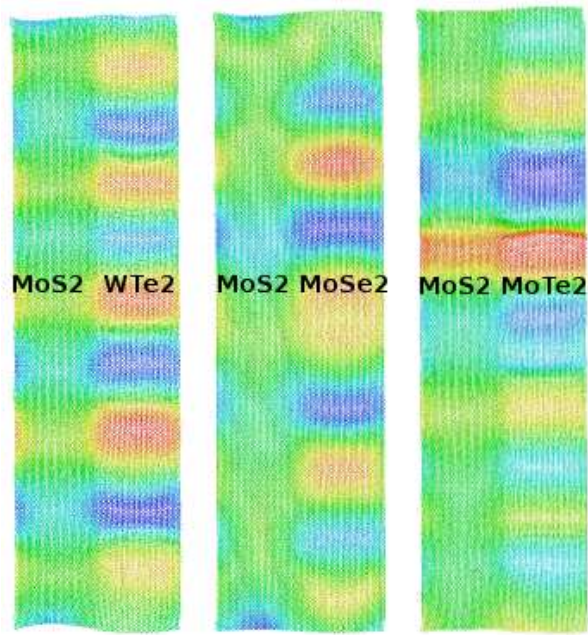}}
  \end{center}
  \caption{(Color online) Misfit strain-induced buckling phenomenon in other TMD lateral heterostructures at 4.2~K. Systems from left to right are: MoS$_2$-WTe$_2$, MoS$_2$-MoSe$_2$, and MoS$_2$-MoTe$_2$. Color represents the atomic z-coordinate.}
  \label{fig_buckling_mx2}
\end{figure}

\begin{figure}[htpb]
  \begin{center}
    \scalebox{1}[1]{\includegraphics[width=8cm]{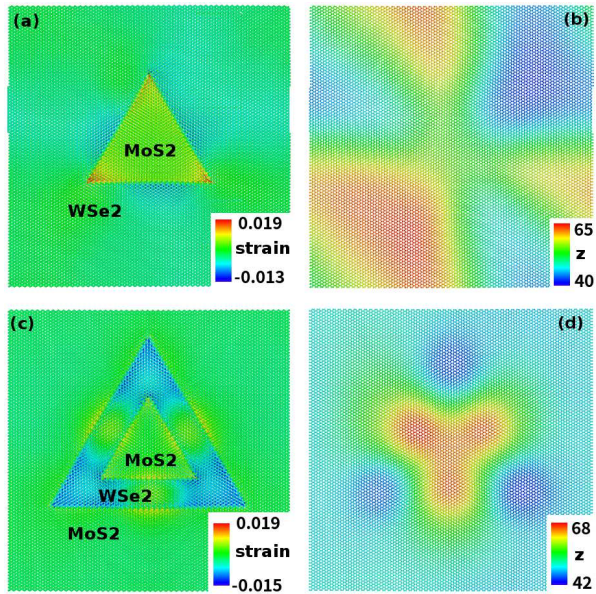}}
  \end{center}
  \caption{(Color online) Misfit strain-induced buckling for TMD lateral heterostructures of triangular pattern at 4.2~K. (a) Strain distribution for MoS$_2$-WSe$_2$ lateral heterostructure, with the buckling shape shown in (b). (c) Strain distribution for MoS$_2$-WSe$_2$-MoS$_2$ lateral heterostructure, with the buckling shape shown in (d). Color corresponds to the strain in (a) and (c); while color represents the z-coordinate in~{\AA} for (b) and (d).}
  \label{fig_triangle_mos2-wse2}
\end{figure}

\begin{figure}[htpb]
  \begin{center}
    \scalebox{1}[1]{\includegraphics[width=8cm]{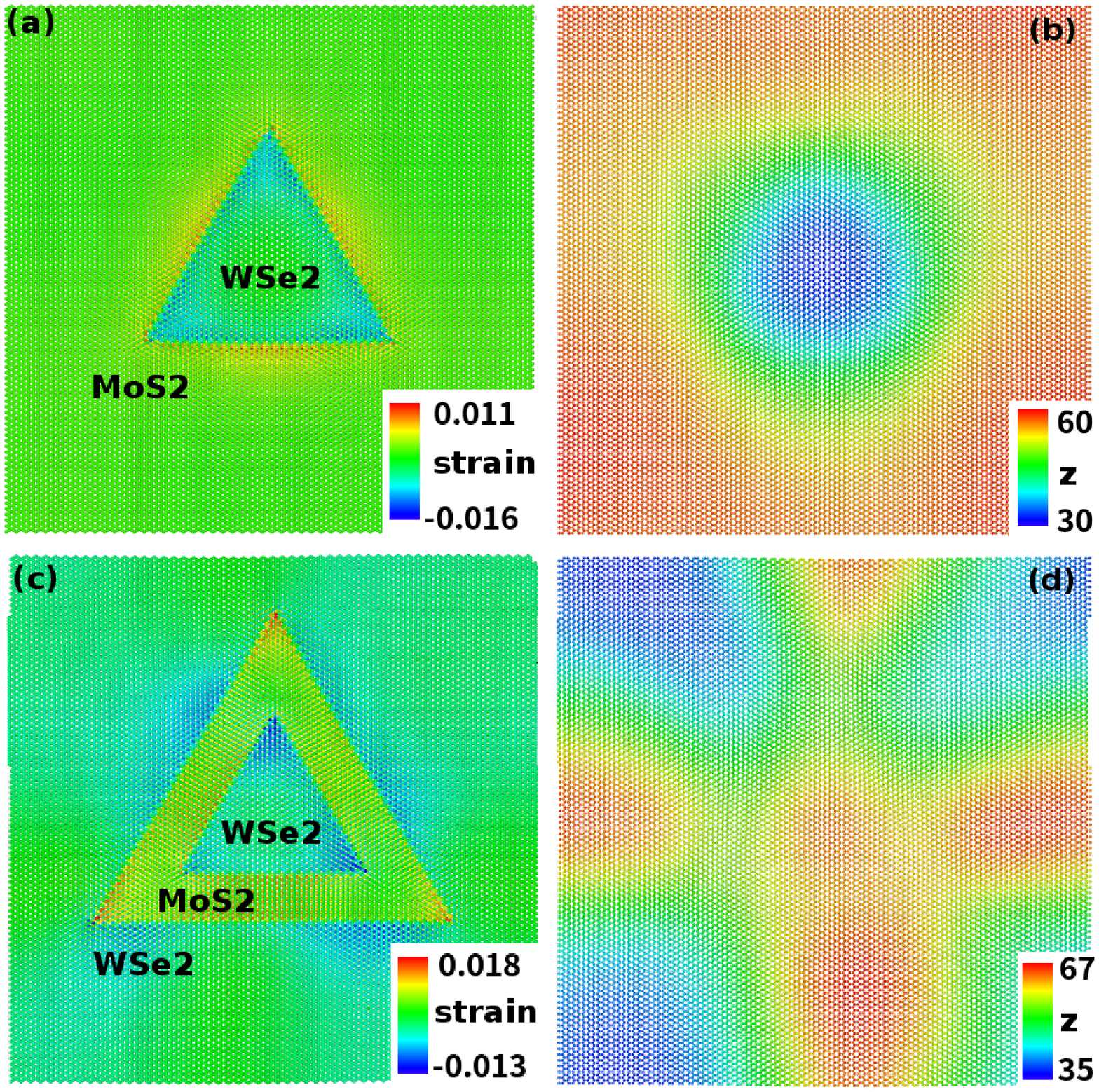}}
  \end{center}
  \caption{(Color online) Misfit strain-induced buckling for the TMD lateral heterostructures of triangular pattern at 4.2~K. The constitutions are switched as compared with the system shown in Fig.~\ref{fig_triangle_mos2-wse2}. (a) Strain distribution for WSe$_2$-MoS$_2$ lateral heterostructure, with the buckling pattern shown in (b). (c) Strain distribution for WSe$_2$-MoS$_2$-WSe$_2$ lateral heterostructure, with the buckling pattern shown in (d). Color corresponds to the strain for (a) and (c); while color represents the z-coordinate in~{\AA} for (b) and (d).}
  \label{fig_triangle_wse2-mos2}
\end{figure}

A particularly interesting result of Fig.~\ref{fig_strain}~(c) is a relatively large compressive strain (-0.8\%) at the WSe$_2$ side of the interface. In contrast to the bulk epitaxial heterostructures, TMDs are atomic-thick with very small bending modulus, so the misfit strain-induced compression effect may induce buckling instability of the TMDs. There is no buckling phenomenon in the MoS$_2$-WSe$_2$ lateral heterostructure with $L_x=200$~{\AA} and $L_y=100$~{\AA} as discussed in the above Figs.~\ref{fig_cfg} and ~\ref{fig_strain}. The misfit strain-induced stress at the interface is along the y-direction, so we study more systems with larger $L_y$. We clearly observe buckling instability for TMD lateral heterostructures of larger $L_y$. Fig.~\ref{fig_buckling_mos2-wse2} illustrates the snapshots for typical buckling shapes of the WSe$_2$ on the right side of the interface for the MoS$_2$-WSe$_2$ lateral heterostructures with $L_y=$ 200, 400, and 800~{\AA}. Fig.~\ref{fig_buckling_amplitude} reveals a critical value of $L_y=105$~{\AA} for the $L_y$ dependence of the ripple amplitude. The ripple amplitude in Fig.~\ref{fig_buckling_amplitude} is estimated by $(z_{\rm max}-z_{\rm min})$, with $z_{\rm max}$ and $z_{\rm min}$ as the maximum and minimum values of the atomic z-coordinate, respectively. For systems with $L_y<105$~{\AA}, the ripple amplitudes are around a small value of 3.5~{\AA} which essentially represents the thickness of the TMD layer. For structures with $L_y>105$~{\AA}, the ripple amplitude is considerably large, reflecting the buckling deformation of the TMD layer. \rev{It has been shown that the buckling of TMD layers will have obvious effects on the electronic transport properties.\cite{KushimaA2015nl} Hence, these buckling patterns induced by the misfit strain can be applied as a valuable approach to tune electronic properties of the TMD lateral heterostructure.}

There is also similar misfit strain-induced buckling instability in other TMD lateral heterostructures. Fig.~\ref{fig_buckling_mx2} shows the optimized configuration of the MoS$_2$-WTe$_2$, MoS$_2$-MoSe$_2$, and MoS$_2$-MoTe$_2$ lateral heterostructures of dimension $L_x\times L_y=200\times 800$~{\AA}. Similar buckling phenomena occur in the TMDs with larger lattice constants, where the misfit induced strain is compressive.

In the above, we have investigated the TMD lateral heterostructures in the rectangular pattern. Experiments have frequently grown the TMD lateral heterostructures of triangular patterns, resulting from the three-fold rotational symmetry of the TMDs. We thus study the misfit strain-induced buckling of the triangular TMD lateral heterostructures. Fig.~\ref{fig_triangle_mos2-wse2}~(a) shows the strain distribution of the triangular MoS$_2$-WSe$_2$ lateral heterostructure, where the MoS$_2$ (WSe$_2$) region is stretched (compressed). There is obvious strain concentration at the three corners of the triangular MoS$_2$ area. Fig.~\ref{fig_triangle_mos2-wse2}~(b) shows the buckling shape of the triangular MoS$_2$-WSe$_2$ lateral heterostructure. The misfit strain-induced buckling also takes place for the triangular MoS$_2$-WSe$_2$ lateral heterostructure with more heteroes as shown in Fig.~\ref{fig_triangle_mos2-wse2}~(c) and (d). Fig.~\ref{fig_triangle_wse2-mos2} shows the misfit strain-induced buckling instability of the triangular WSe$_2$-MoS$_2$ lateral heterostructure, where the sequence of MoS$_2$ and WSe$_2$ is exchanged as compared with the structure shown in Fig.~\ref{fig_triangle_mos2-wse2}.

\section{conclusion}
In conclusion, we have performed molecular dynamics simulations to study the misfit strain-induced buckling phenomena in the MoS$_2$-WSe$_2$, MoS$_2$-WTe$_2$, MoS$_2$-MoSe$_2$, and MoS$_2$-MoTe$_2$ lateral heterostructures, where the atomic interactions for these multiple-component systems are described by the SW potential. We explored the strain distribution within the heterostructure, which is consistent with the experiment. We found that the misfit strain can cause buckling of the TMD with larger lattice constants in the lateral heterostructure, due to the atomic-thick nature of TMDs. The misfit strain-induced buckling is quite general for TMD lateral heterostructures of different constitutions and different hetero-patterns.

\textbf{Acknowledgements} The work is supported by the Recruitment Program of Global Youth Experts of China, the National Natural Science Foundation of China (NSFC) under Grant No. 11504225, and the Innovation Program of Shanghai Municipal Education Commission under Grant No. 2017-01-07-00-09-E00019.

\appendix
\section{Modification for LAMMPS}

We point out one necessary modification for the three-body SW potential implemented in LAMMPS. More details on this modification can be found from our earlier work.\cite{JiangJW2016swborophene} Overall, the modification is done in two steps.

(1). First, find the following line in the {pair\_sw.cpp} source file,

\hspace{1cm} delcs = cs - paramijk-$>$costheta;

(2). Second, insert the following new lines after the above line,

\hspace{1cm}  if(fabs(delcs) $<=$ 0.25)

\hspace{1cm}  \{

\hspace{1.5cm}     delcs = delcs;

\hspace{1cm}  \}

\hspace{1cm}  else if(fabs(delcs) $<$ 0.35)

\hspace{1cm}  \{

\hspace{1.5cm}     delcs = delcs * (0.5+0.5*sin(3.142*(delcs-0.25)/(0.35-0.25)+0.5*3.142));

\hspace{1cm}  \}

\hspace{1cm}  else

\hspace{1cm}  \{

\hspace{1.5cm}     delcs = 0.0;

\hspace{1cm}  \}
\\Then recompile the LAMMPS package. The recompiled LAMMPS executable file can be used to simulate materials with inequivalent angles around each atom (like TMDs in the present work) using the SW potential. This modification does not affect other simulations with LAMMPS.

%
\end{document}